\documentclass[aps,prl,reprint,superscriptaddress,showpacs,longbibliography,floatfix,footnoteinbib]{revtex4-1}
\usepackage{amsmath,amssymb,graphicx,units}
\usepackage[plainpages=false,pdfpagelabels,colorlinks=true,linkcolor=red,urlcolor=blue,citecolor=blue,pdftitle={Title},pdfauthor={},pdfdisplaydoctitle=true,pdfduplex=DuplexFlipLongEdge]{hyperref}

\newcommand{\beginsupplement}{%
        \setcounter{table}{0}
        \renewcommand{\thetable}{S\arabic{table}}%
        \setcounter{figure}{0}
        \renewcommand{\thefigure}{S\arabic{figure}}%
}

\begin{document}

\title{Giant Magnetoresistance in Hubbard Chains}

\author{Jian Li}
\author{Chen Cheng}
\email{chengchen@csrc.ac.cn}
\affiliation{Beijing Computational Science Research Center, Beijing 100193, China}
\author{Thereza Paiva}
\affiliation{Instituto de F\'\i sica, Universidade Federal do Rio de Janeiro Cx.P. 68.528, 21941-972 Rio de Janeiro RJ, Brazil}
\author{Hai-Qing Lin}
\author{Rubem Mondaini}
\email{rmondaini@csrc.ac.cn}
\affiliation{Beijing Computational Science Research Center, Beijing 100193, China}
\begin{abstract}
We use numerically unbiased methods to show that the one-dimensional  Hubbard model with periodically distributed on-site interactions already contains the minimal ingredients to display the phenomenon of magnetoresistance; i.e., by applying an external magnetic field, a dramatic enhancement on the charge transport is achieved. We reach this conclusion based on the computation of the Drude weight and of the single-particle density of states, applying twisted boundary condition averaging to reduce finite-size effects. The known picture that describes the giant magnetoresistance, by interpreting the scattering amplitudes of parallel or antiparallel polarized currents with local magnetizations, is obtained without having to resort to different entities; itinerant and localized charges are indistinguishable.
\end{abstract}


\maketitle

\paragraph{Introduction.---}

The phenomenon of giant magnetoresistance highlights the speed in which some results in basic research can be rapidly converted into technological applications. It took less than a decade from its discovery in the late 1980s~\cite{Baibich1988, Binasch1989} to its implementation on the read heads of high-density hard disks commercialized for the general public. Specifically, it describes the significant reduction of electrical resistance of certain materials, composed of sandwiches of thin magnetic and nonmagnetic layers, in the presence of an external magnetic field. The physical explanation of this purely quantum mechanical effect relies on the fact that electrons traveling through a ferromagnetic conductor will scatter differently depending on the relative orientation of their spin to the magnetization direction of the conductor---with those oriented parallel scattering less often than those oriented antiparallel~\cite{Gerstner2007,Grunberg2008,Nobel}.

\begin{figure}[b!] 
\includegraphics[width=0.99\columnwidth]{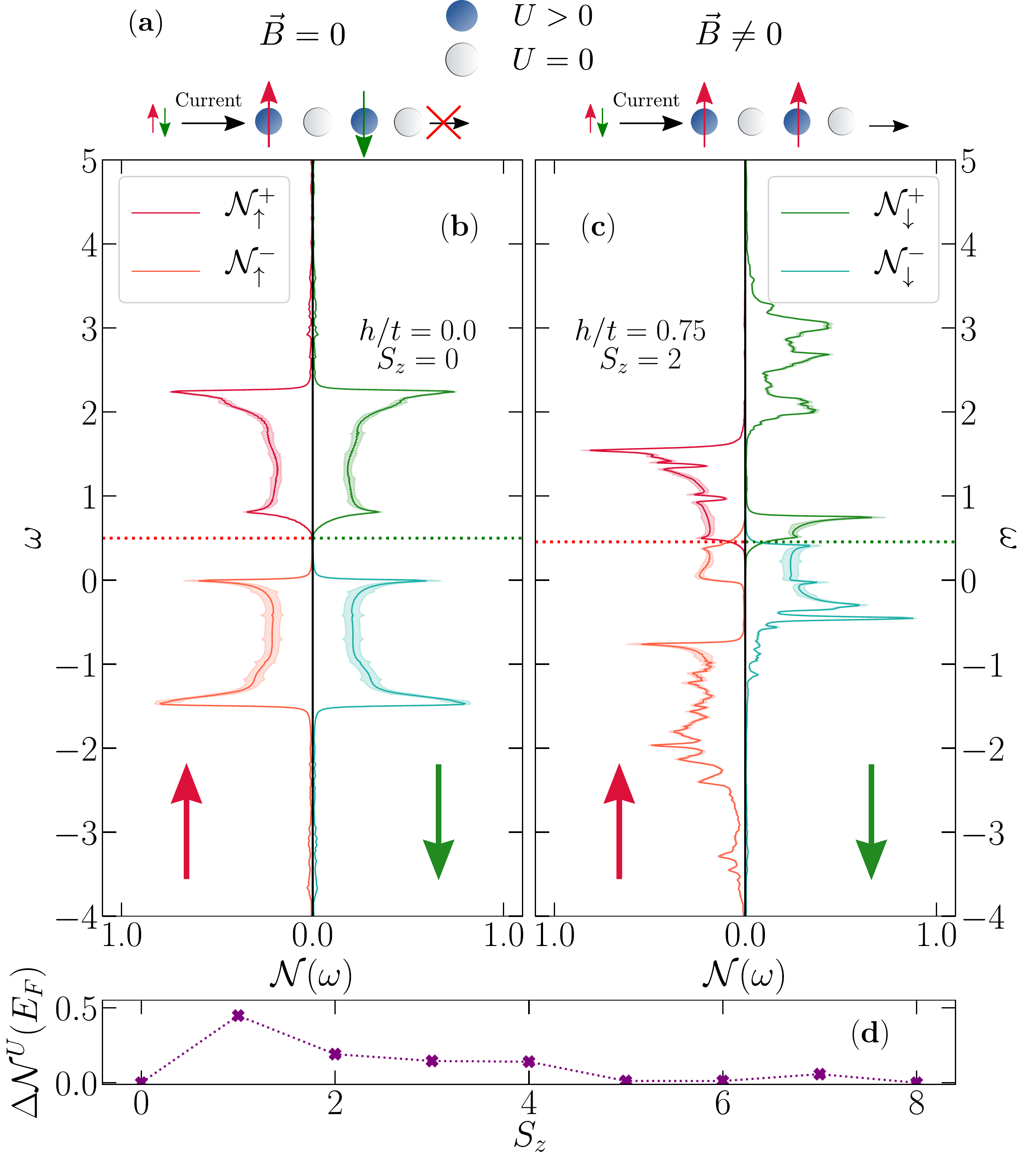}
\caption{(a) Schematic representation of the superlattice with the picture for transport and magnetism (see the text), in the absence or presence of an external magnetic field $\vec B$. In (b) [(c)], we display the spin-resolved density of states of the superlattice ($L=16$) at zero temperature for $h = 0$ [$h\neq 0$]. In the absence of the field, the Mott gap renders an insulating behavior, while the latter, a metal induced by the field, has a much higher mobility for charges with spin aligned to $\vec B$, as highlighted by the difference in local density of states at the Fermi energy for finite population imbalances in (d). Shading surrounding the curves depicts the error bars after the twisted boundary condition averaging and dashed lines, the Fermi energies.}
 \label{fig:dos_final}
\end{figure}

In a band picture, this is explained by the imbalance of charge populations with spin parallel and antiparallel to an external magnetic field, which translates into very different local density of states in the magnetic regions for both spin states at the Fermi energy~\cite{Nobel}. For the antiparallel component, the reduced density of states results in a higher resistance for this channel, compared to a lower resistance for the parallel one. A simplified model of resistances based on the scattering of each itinerant spin component by the magnetization of the background, qualitatively explains the increased conductivity in these materials, since the external field polarizes the magnetization of the ferromagnetic layers, and it thus enhance the transport for electrons that have spin parallel to it~\cite{Zutic2004,Gerstner2007,Grunberg2008,Nobel}.

Interestingly, this has also been investigated within the scope of \textit{ab initio} electronic structure calculations, which do not account, \textit{per se}, for interactions between electrons -- but are complemented by spin-dependent scattering using quasiclassical methods~\cite{Zahn1995,Richter1996}. Here, our approach is different: We start from the simplest possible interacting model describing electrons hopping on a lattice with reduced dimensionality -- essentially a one-dimensional chain or a nanowire -- and model magnetic and nonmagnetic regions via site-dependent (although periodic) interactions. By unbiasedly calculating the transport properties of this simplified system, we show that it already contains the necessary attributes to display effects similar to the giant magnetoresistance (GMR) phenomenon in a purely interacting setting, as schematically represented in the sketch in Fig.~\ref{fig:dos_final}. Besides, what is mostly considered a phenomenon that arises from the interplay of two types of electrons, localized and delocalized ones, here is obtained via a single entity.

\paragraph{Model and methods. ---}
We use the one-dimensional Hubbard model with site-dependent interactions~\cite{Paiva1996,Paiva1998,Paiva2000,Paiva2002,Malvezzi2002,Zhang2015}, creating a superlattice of size $L$,
\begin{eqnarray}
  \mathcal{\hat H} &=&  -t \sum_{i,\sigma} \left(\hat{c}_{i,\sigma}^{\dagger} \hat{c}_{i+1,\sigma} + {\rm H.c.} \right)  + \sum_{i} U_i \hat{n}_{i,\uparrow} \hat{n}_{i,\downarrow} \nonumber \\
  && -h \sum_{i}\left(\hat{n}_{i,\uparrow} - \hat{n}_{i,\downarrow}\right),
  \label{eq:Hamiltonian}
\end{eqnarray}
where $\hat{c}_{i,\sigma}^\dagger$($\hat{c}_{i,\sigma}$) creates(annihilates) a fermion with spin $\sigma$ ($\uparrow$ or $\downarrow$) at the $i$th site of the lattice,  and $\hat{n}_{i,\sigma} = \hat{c}_{i,\sigma}^{\dag} \hat{c}_{i,\sigma}$. The first term in~(\ref{eq:Hamiltonian}) accounts for the hopping of electrons between nearest-neighbor sites; $U_i$ is the on-site Coulomb repulsion energy and $h$ is the Zeeman energy related to an applied magnetic field $\vec B$. $t$ sets the energy scale of the problem; we assume cyclic boundary conditions, and restrict our results to half filling ($\sum_{i,\sigma} \langle \hat n_{i,\sigma}\rangle/L = 1$; $\langle \cdot \rangle$ is understood as the ground state average). For the interactions, we focus on the case where they are chosen in a periodic fashion with the repeated intercalation of the $U_i = U>0$ and $U_i = 0$ sites [Fig.~\ref{fig:dos_final}(a)].

\begin{figure}[h] 
\includegraphics[width=0.99\columnwidth]{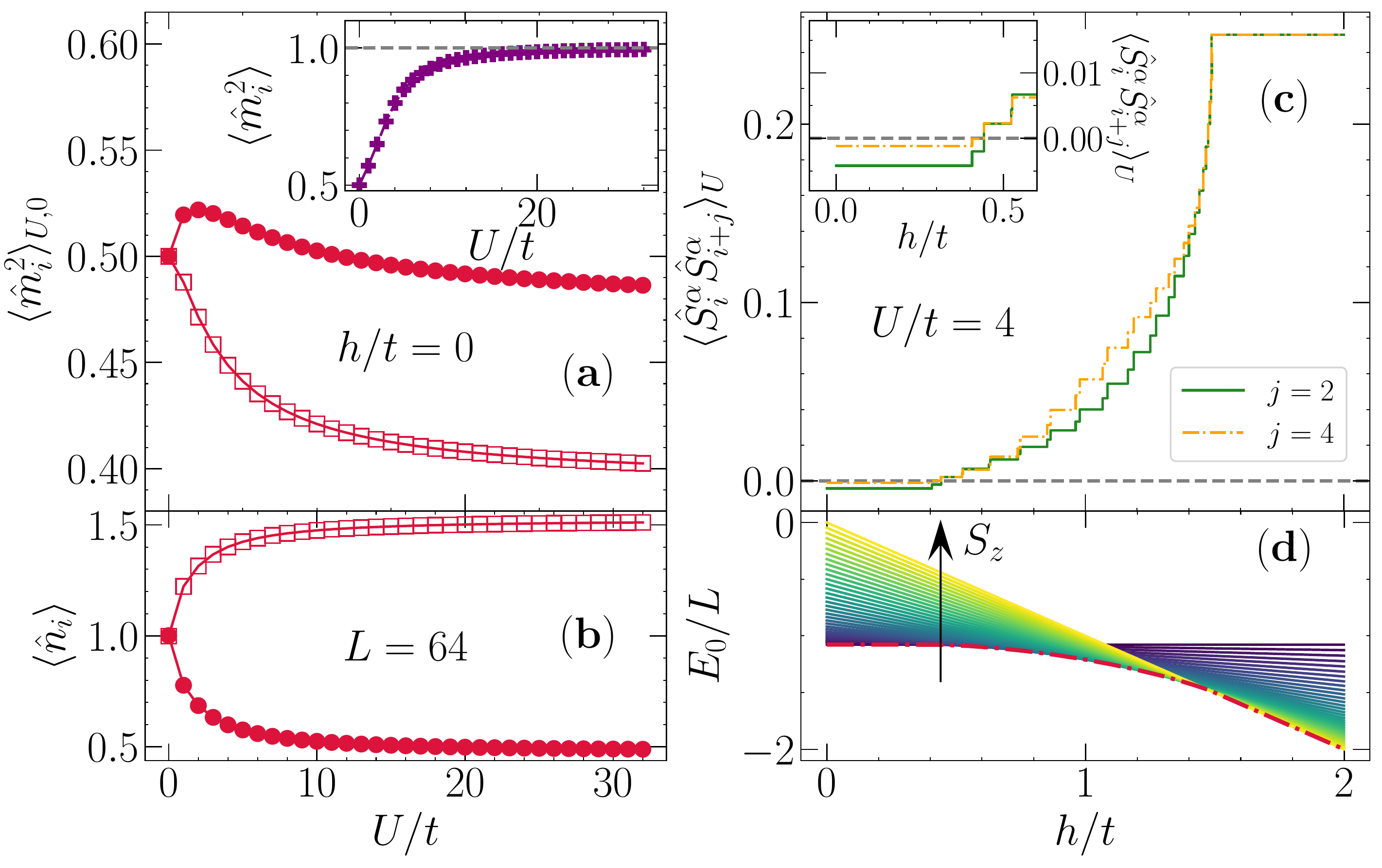}
\caption{Local moment $\langle \hat m_i^2\rangle_U$ ($\langle \hat m_i^2\rangle_0$) and its dependence on the interaction strength in repulsive (free) sites, marked by full (empty) symbols in (a), for the superlattice with $h/t=0$. (Inset) The same for a homogeneous lattice; the dashed line denotes the Heisenberg limit of full localization. Site-dependent interactions break particle-hole symmetry and lead to an imbalance of the densities in both types of sites, as shown in (b). In (c), the spin correlations for nearest and next-nearest repulsive sites, taking a repulsive site as the reference, as a function of the external field magnitude with $U/t=4$. (Inset) The negative correlations for $h=0$. All results are presented for a lattice with $L=64$, using DMRG. (d) Dependence on the Zeeman field of the lowest energy state for different $S_z$ sectors of Eq.~\eqref{eq:Hamiltonian}. The ground state is represented by the lower dashed-dotted curve enveloping the lines of each sector.}
 \label{fig:correlations}
\end{figure}

To understand how this simple model leads to a crude interpretation of magnetic and nonmagnetic regions, it is useful to recall the dependence of the local moment, $\langle \hat m_i^2\rangle \equiv \langle (\hat n_{i,\uparrow} - \hat n_{i,\downarrow})^2\rangle$, on the interaction magnitude within a homogeneous lattice [inset of Fig.~\ref{fig:correlations}(a)]. Starting from the noninteracting regime, it assumes a value of 1/2, at this density, and steadily increases towards 1, when approaching the Heisenberg limit for large $U$. Thus, interactions induce the formation of magnetic moments, and when generalizing to a superlattice configuration, this is still the case, albeit less dramatically due to a natural density imbalance between repulsive and free sites [Figs.~\ref{fig:correlations}(a) and~\ref{fig:correlations}(b)]. This argument leads to the simple association that $U>0$ and $U=0$ types of sites can mimic the physics of magnetic and nonmagnetic regions in actual materials. 

In what follows, we have used Lanczos diagonalization~\cite{Lin1993} and density matrix renormalization group (DMRG)~\cite{White1992,White1993} to obtain the ground state properties of the superlattices. We notice that when dealing with independent sectors of the Hamiltonian with a given total magnetization in the $z$ direction, $S_z = \frac{1}{2}\left(N_\uparrow - N_{\downarrow}\right)$ [$N_\sigma$ is the total number of particles with spin $\sigma$], the Zeeman energy is trivially accounted for and results in a shift of the energies for finite values of the external magnetic field. Thus, as $h$ grows, different sectors will host the ground state of the Hamiltonian, as exemplified in Fig.~\ref{fig:correlations}(d).

\paragraph{Density of states. --- } 
To see how this space-dependent local moment affects the transport properties in Eq.~\eqref{eq:Hamiltonian} and connects our problem to the known phenomenology of the GMR effect, we obtain the density of states by computing single-particle excitations in the ground state. This is accomplished by numerically calculating dynamical quantities as the spectral function~\cite{Gagliano1987,Dagotto1994}, $A_k^\sigma(\omega) = \sum_n |\langle \psi_0| \hat c^\dagger_{k,\sigma}| \psi_n^{N_\sigma-1}\rangle|^2 \delta\left(\omega + (E_n^{N_\sigma-1}-E_0)\right) + \sum_n |\langle \psi_0| \hat c_{k,\sigma}| \psi_n^{N_\sigma+1}\rangle|^2 \delta\left(\omega - (E_n^{N_\sigma+1}-E_0)\right)$, which describes the dynamical response of creating a fermion and a hole with momentum $k$ and spin $\sigma$ in the ground state $|\psi_0\rangle$ (with eigenenergy $E_0$) of the Hamiltonian; $|\psi_n^{N_\sigma\pm1}\rangle$ ($E_n^{N_\sigma\pm1}$) are eigenstates (eigenvalues) of the Hamiltonian with an added or removed electron. When summing up all possible momentum excitations, one recovers the actual density of states, ${\cal N}_\sigma(\omega) = (2/L)\sum_k A_k^\sigma(\omega) = {\cal N}^+_\sigma(\omega) + {\cal N}^-_\sigma(\omega)$, where we have resolved the contributions for electron and hole excitations in the last equality. To mitigate the influence of finite-size errors, we have employed twisted boundary condition averaging~\cite{Gammel1992,Gros1996, SM}; this has been used in a variety of contexts so as to approach the results in the thermodynamical limit with limited system sizes~\cite{Chiesa2008,Varney2011,MendesSantos2015,Schuetrumpf2015,Qin2016}, and it has been shown~\cite{Poilblanc1991} to be especially relevant for the case of dynamical quantities~\cite{SM}.

In Figs.~\ref{fig:dos_final}(b) and~\ref{fig:dos_final}(c), we report this quantity for $h = 0$ and $h\neq0$, respectively, for $U/t=4$, averaged among 64 boundary conditions. In the absence of an external field, the interactions, even if not present in every site, induce the formation of a Mott gap separating the lower and upper Hubbard bands; therefore, the ground state is a perfect insulator. Now, by applying an external magnetic field, the ground state no longer has the total $S_z=0$, but, rather, finite values. For, say $h/t=0.75$, single-particle excitations in the ground state, which has $S_z=2$, display a \textit{metallic} behavior~\footnote{The Fermi energies can be seen to be at the intersection of the single-particle excitation types, ${\cal N}^+_\sigma(\omega)$ and ${\cal N}^-_\sigma(\omega)$, and are verified through the calculation of the total number of particles for each spin species via integration of the spin-resolved density of states.}. Moreover, the difference in the local density of states of both spin channels in repulsive sites at the Fermi energy, $\Delta {\cal N}^{U} = {\cal N}_{\uparrow}^{U}-{\cal N}_{\downarrow}^{U}$~\cite{SM}, shows that the transport is facilitated  when there is a population imbalance [Fig.~\ref{fig:dos_final}(d)]. Hence, if one injects a non-spin-polarized current in the superlattice [see Fig.~\ref{fig:dos_final}(a)], the transport is enhanced, similar to the GMR effect, also realized in nanowires~\cite{Piraux1994,Blondel1994,Liu1995}. Now, this is one of the differences between the standard GMR and our results: In the actual experiments, the material, being metallic, possesses a finite conductivity which is enhanced by the application of a magnetic field. Here, we start from a perfect insulator and see that it induces metallic behavior. In other words, the model we investigate displays \textit{perfect} magnetoresistance, provided the field is sufficiently large to induce a finite magnetization in the ground state.

\paragraph{Relative magnetization. ---} A further characterization of the similarity between our results and the GMR physics, can be seen through spin correlations. We notice that in the latter, transport is enhanced when the magnetization of consecutive ferromagnetic layers is made parallel. In Fig.~\ref{fig:correlations}(c), we show the dependence on the Zeeman energy of the spin-spin correlation $\langle \hat S_i^\alpha \hat S_{i+j}^\alpha\rangle_U \equiv (1/4)\langle m_i^\alpha m_{i+j}^\alpha\rangle_U$, where $i$ is a repulsive site and $j$ is either the nearest or next-nearest site, also with $U>0$; $\alpha$ is the direction of the applied Zeeman field. We notice that for the values of the field where we observe the enhancement on the transport via the analysis of $N(\omega)$, these spin correlations are positive, denoting parallel orientation, while they are slightly negative in its absence. The arrows in Fig.~\ref{fig:dos_final}(a) schematically represent this situation.

\paragraph{Transport properties -- Drude weight. --- } A robust way of checking the transport properties of quantum systems is via the Drude weight, $D/\pi e^2$, that measures the density of mobile charge carriers to their mass, or charge stiffness~\cite{Fye1991,Scalapino1993}, {i.e.}, in the thermodynamic limit $D\neq0$ ($D=0$) signals a metallic (insulating) behavior. This quantity appears in the real part of the $q=0$ optical conductivity, $\sigma(\omega) = D\delta(\omega) + \sigma_{\rm reg}(\omega)$, as a weight for the singular behavior at zero frequency; it has also been shown by Kohn~\cite{Kohn1964} that it can be computed from the change of the ground state energy $E_0$ to an applied flux $\Phi$ on the lattice as~\cite{Giamarchi2004},
\begin{equation}
\frac{D}{\pi e^2} = L \left(\frac{\partial^2 E_0}{\partial \Phi^2}\right) \Big\vert_{\Phi=0},
\label{eq:drude}
\end{equation}
being related to the induction of persistent currents in the system. The flux is introduced in the Hamiltonian (\ref{eq:Hamiltonian}) via a Peierls substitution on the hopping terms of the Hamiltonian, i.e., $-t \hat{c}_{i,\sigma}^{\dagger} \hat{c}_{i+1,\sigma} \to -t e^{{\rm i} \phi} \hat{c}_{i,\sigma}^{\dagger} \hat{c}_{i+1\sigma}$~\cite{Stafford1991,Fye1991,Scalapino1993}, where $\phi = \Phi/L$~\footnote{We use Peierls phases spread over each bond of the lattice so as to preserve translational symmetry, which is accounted for in the basis formation when using Lanczos ground state diagonalization. A trivial gauge transformation relates this case to the one with a single phase at the ends of the chain.}. It is important to highlight that these phases are of merely mathematical help and do not alter the external magnetic field introduced in the Zeeman term of Eq.~\eqref{eq:Hamiltonian}, since the latter could be taken as perpendicular to the field associated to the flux $\Phi$. Besides, they also do not change physical observables, as, e.g., densities~\cite{Giamarchi2004}

\begin{figure}[t] 
\includegraphics[width=0.99\columnwidth]{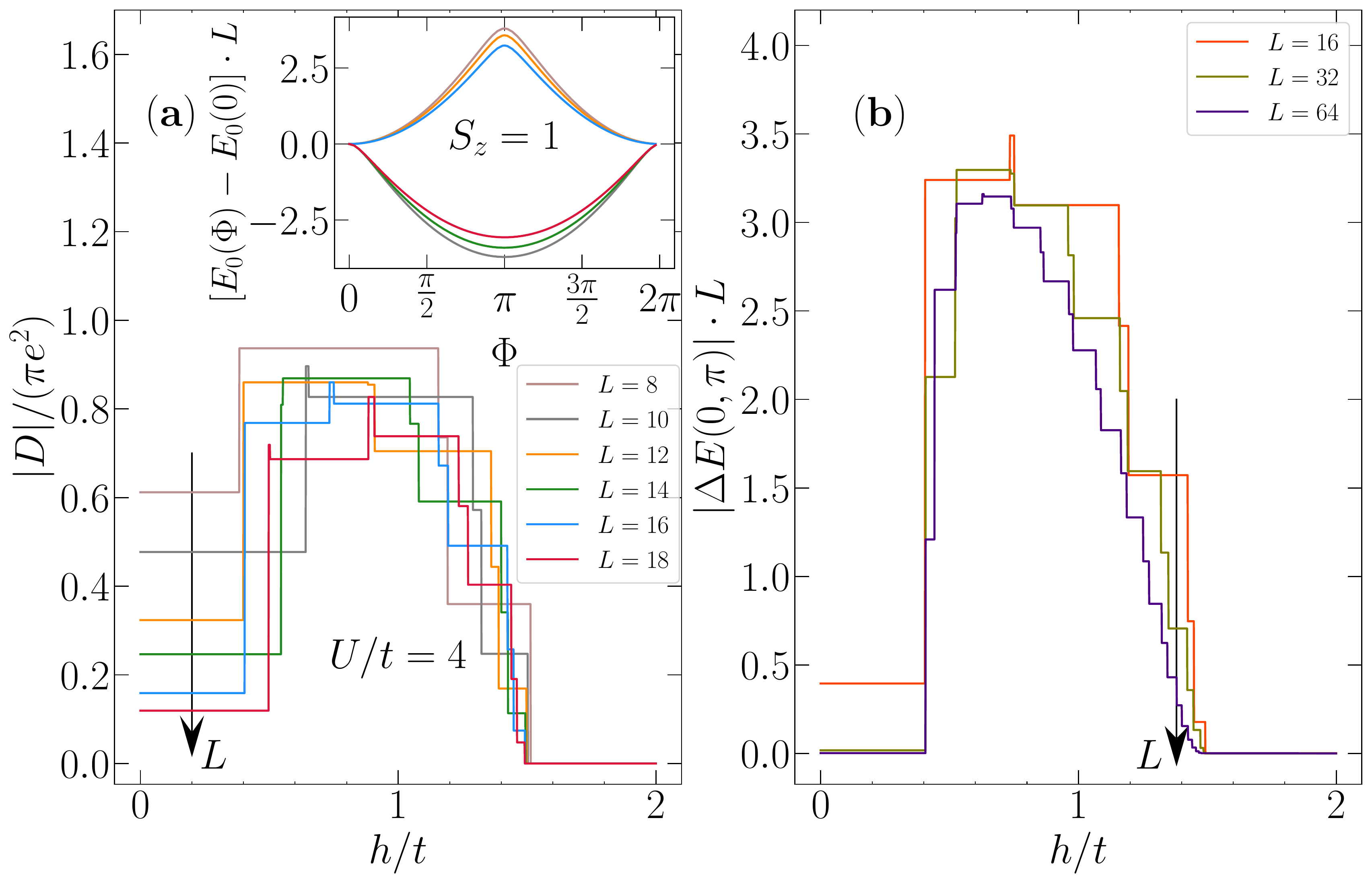}
\caption{ (a) Drude weight dependence on the Zeeman energy for the superlattice with interaction strength $U/t = 4$ and different system sizes. These are obtained via Lanczos diagonalization after using Eq.(\ref{eq:drude}) to obtain the curvatures of $E_0 L$ vs $\Phi$ curves; an example for zero field is presented in the inset. (b) The energy difference between periodic and antiperiodic boundary conditions for much larger system sizes obtained via DMRG as a function of the Zeeman energy.}
 \label{fig:Drude_and_DeltaE_vs_hmag}
\end{figure}

A typical dependence of the ground state energy of the superlattice with the flux $\Phi$, for $S_z=1$, is presented in the inset of Fig.~\ref{fig:Drude_and_DeltaE_vs_hmag}(a), for different values of $L$. Lattices with $L=4n$ ($n$ is an integer) are known to display a paramagnetic response ($D < 0$)~\cite{Fye1991}. For that reason, we focus on the absolute values of $D$ and its dependence on the Zeeman field, in Fig.~\ref{fig:Drude_and_DeltaE_vs_hmag}(a), for different system sizes to understand whether it can show signatures of the enhancement of transport as observed in the density of states. Likewise, Fig.~\ref{fig:Drude_and_DeltaE_vs_hmag}(b) shows the corresponding difference in energy between the cases with periodic $(\Phi = 0)$ and antiperiodic boundary conditions $(\Phi = \pi)$, $\Delta E(0,\pi)$, by using DMRG calculations in much larger lattices. Since the difference in energies will be finite as long as the curvature of $E_0(\Phi)$ at $\Phi=0$ is finite, provided there are no other local minima or maxima in $0<\Phi<\pi$, it is suitable to track $\Delta E(0,\pi)$, as one deals only with real numbers in the numerics. The qualitative behavior for the two quantities is similar: An initially  finite and small Drude weight is suddenly increased after the ground state acquires a finite magnetization, for growing values of the field. At an even larger $h/t$, the transport decreases and the system becomes (band) insulating at a saturation Zeeman energy $h_{\rm sat}$. This corresponds to the situation where the ground state is fully polarized and the Pauli exclusion principle prevents any charge mobility.

\begin{figure}[t] 
\includegraphics[width=0.99\columnwidth]{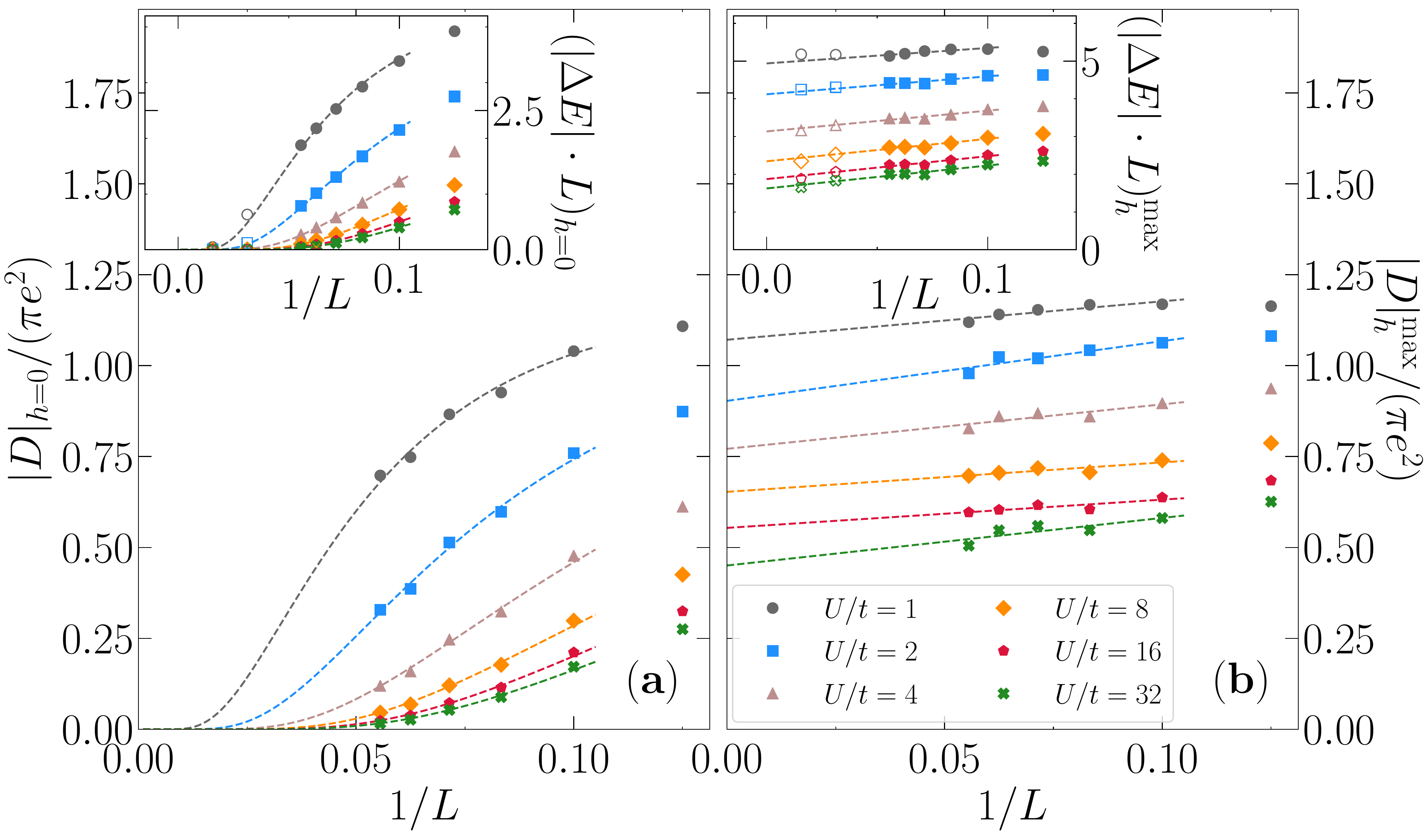}
\caption{System size scaling of the Drude weight in (a) the absence of magnetic field and (b) for the value of $h$ that results in the largest $D$ for a given system size. (Insets) The respective scaling analyses for $\Delta E(0,\pi)$, where the empty symbols denote the DMRG results, for larger systems.}
 \label{fig:scaling_Drude}
\end{figure}

A finite-size scaling is in order to assess the thermodynamic limit. We report in Fig.~\ref{fig:scaling_Drude}(a) the system size dependence of $D$ at $h=0$ and at the value of the field that gives the maximum Drude weight, $|D|_{h}^{\rm max}$; the insets display the same for $\Delta E(0,\pi)$ (with qualitative similar results), comparing a wide range of interactions $U/t$. In the former, we notice that by using the functional form of Ref.~\cite{Stafford1991}, $|D|\propto \sqrt{L/\xi}e^{-L/\xi}$ ($\xi$ is the Mott localization length), derived from the Bethe ansatz equations and thus valid for homogeneous chains, one can equally fit our data in the case of superlattices. Remarkably, half-filled superlattices possess insulating behavior when $L\to\infty$ in the absence of an external magnetic field, i.e., $D_{L\to\infty}\to0$. On the other hand, for the maximum Drude weight, a linear extrapolation with $1/L$ results in finite $D$ values [or $\Delta E(0,\pi)$]: The introduction of a magnetic field induces transport of the charges or, more precisely, an insulator-to-metal transition, for $h/t\approx 0.5$, and is particular to superlattices~\cite{SM}.

\begin{figure}[t] 
\includegraphics[width=0.9\columnwidth]{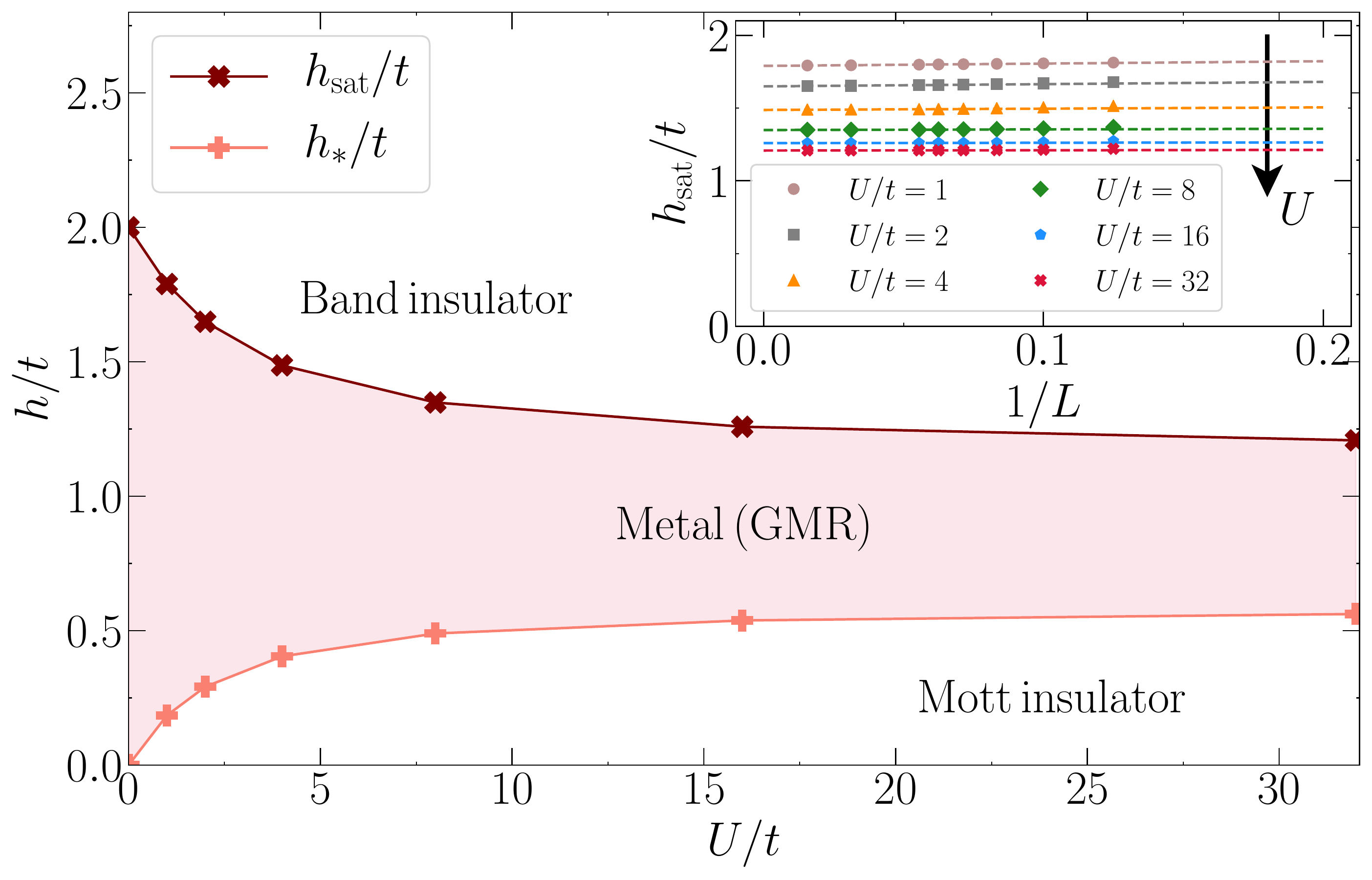}
\caption{Regime of parameters where the insulator-to-metal transition is observed (shaded area). Small (large) values of the field, result in a Mott (band) insulator. (Inset) The finite-size scaling of the saturation field, shown as an example. $h_{\rm sat}$ can be similarly obtained via an analysis of an effective two-body problem~\cite{SM}.}
 \label{fig:GMR_map}
\end{figure}

This is valid in the regime where $U/t$ is finite -- since increasing the interactions leads to a smaller Drude weight in large lattices [Fig.~\ref{fig:scaling_Drude}(b)]. Apart from that, the enhancement of $D$ in respect to the insulating case is constrained to regimes of finite magnetizations of the ground state other than $S_z = N_\uparrow/2=L/2$. This generates a range of values of $h/t$ where the magnetoresistance in our model can be manifest. Figure~\ref{fig:GMR_map} analyzes how this range depends on the interaction magnitudes, being limited by $h_*$, where the ground state no longer has $S_z=0$, and $h_{\rm sat}$.

\paragraph{Summary and discussion. --- }
We used a simple model, the Hubbard model with periodic site-dependent interactions under the presence of an external magnetic field, and we  identify results analogous to the GMR phenomenon in a purely interacting setting. This is achieved via the identification of repulsive (noninteracting) regions as being magnetic (nonmagnetic), similar to the `sandwiches' of ferromagnetic and nonmagnetic layers in experimental samples. The combined quantification of transport and spin correlation functions show that when the magnetization in consecutive ``magnetic'' regions is made parallel due to the application of the magnetic field, the transport is enhanced, and an insulator-to-metal transition is obtained. An investigation of other densities and configurations of the superlattices may be relevant in the optimization of these features but goes beyond the scope of this Letter. 

Most importantly, these results transcend the curiosity of solving a simple interacting model and have the possibility of being emulated using cold atoms trapped in optical lattices; charges and spin degrees of freedom are then translated into atoms and its hyperfine states, respectively. Besides, spatially dependent interactions are becoming a reality in experiments of ultracold gases. The usage of optical control to induce Feshbach resonances~\cite{Bauer2009,Yamazaki2010,Wu2012,Wu2012b,Jagannathan2016} and, consequently, local interactions, has witnessed new breakthroughs~\cite{Clark2015,Arunkumar2018}, that we envision being sufficient to investigate the space-dependent interactions of this model. Last, a verification of our results in experiments would require a precise quantification of transport of trapped atoms. Recently, however, this has been shown to be achievable when emulating the Hubbard model, either when focusing on spin~\cite{Nichols2018} or charge~\cite{Brown2018} degrees of freedom. For this reason, our results may inspire experimentalists in understanding this highly unusual transport mechanism, which has a deep connection with the GMR effect, a phenomenon usually constrained to the condensed matter realm.

\begin{acknowledgments}
  The authors acknowledge insightful discussions with R.R.~dos Santos, P.D.~Sacramento, G.~Batrouni, R.T.~Scalettar and S.-W. Tsai. We acknowledge support by NSFC Grants No. U1530401 (J.L., C.C., R.M. and H.Q.L.), No. 11674021 (C.C. and R.M.) and No. 11650110441 (R.M.); T.P. acknowledges the financial support of CNPq and INCT on Quantum Information. The computations were performed in the Tianhe-2JK at the Beijing Computational Science Research Center (CSRC).
\end{acknowledgments}

\bibliography{references}

\clearpage

\onecolumngrid

\begin{center}

{\large \bf Supplementary Materials:
\\ Giant Magnetorresistance in Hubbard Chains}\\

\vspace{0.3cm}

\end{center}

\vspace{0.6cm}

\twocolumngrid

\beginsupplement

\section{Twisted boundary conditions averaging and the density of states}
\subsection{Non-interacting regime}\label{sec:nonint}

Twisted boundary conditions (TBC) have been extensively used as a way to reduce finite size effects on observables of tight-binding systems for a long time. It allows one to effectively  investigate other $k$-points, which are not originally manifest in a finite system size, with the goal of better capturing the physics when approaching the thermodynamic limit. For example, in a non-interacting tight-binding 1D system, the dispersion relation for the fermions in the case of periodic boundary conditions, is given by $\varepsilon(k) = -2t\cos\left (k\right)$, with $k = \frac{2n\pi}{L}$ and $n\in \left[-\frac{L}{2},\frac{L}{2}\right)$. After introducing the Peierls phases, $c^\dagger_j\to c^\dagger_j e^{{\mathrm i} \phi j}$, where $\phi = \Phi/L$, the non-interacting single-particle spectrum is then modified to $\varepsilon(k, \Phi) = -2t\cos\left( k + \Phi/L\right)$. Thus, for each configuration of those phases, it allows one to probe $k$-points that would only be present on much larger lattices with standard periodic boundary conditions.

\begin{figure}[t]
  \centering
  \includegraphics[width=0.99\columnwidth]{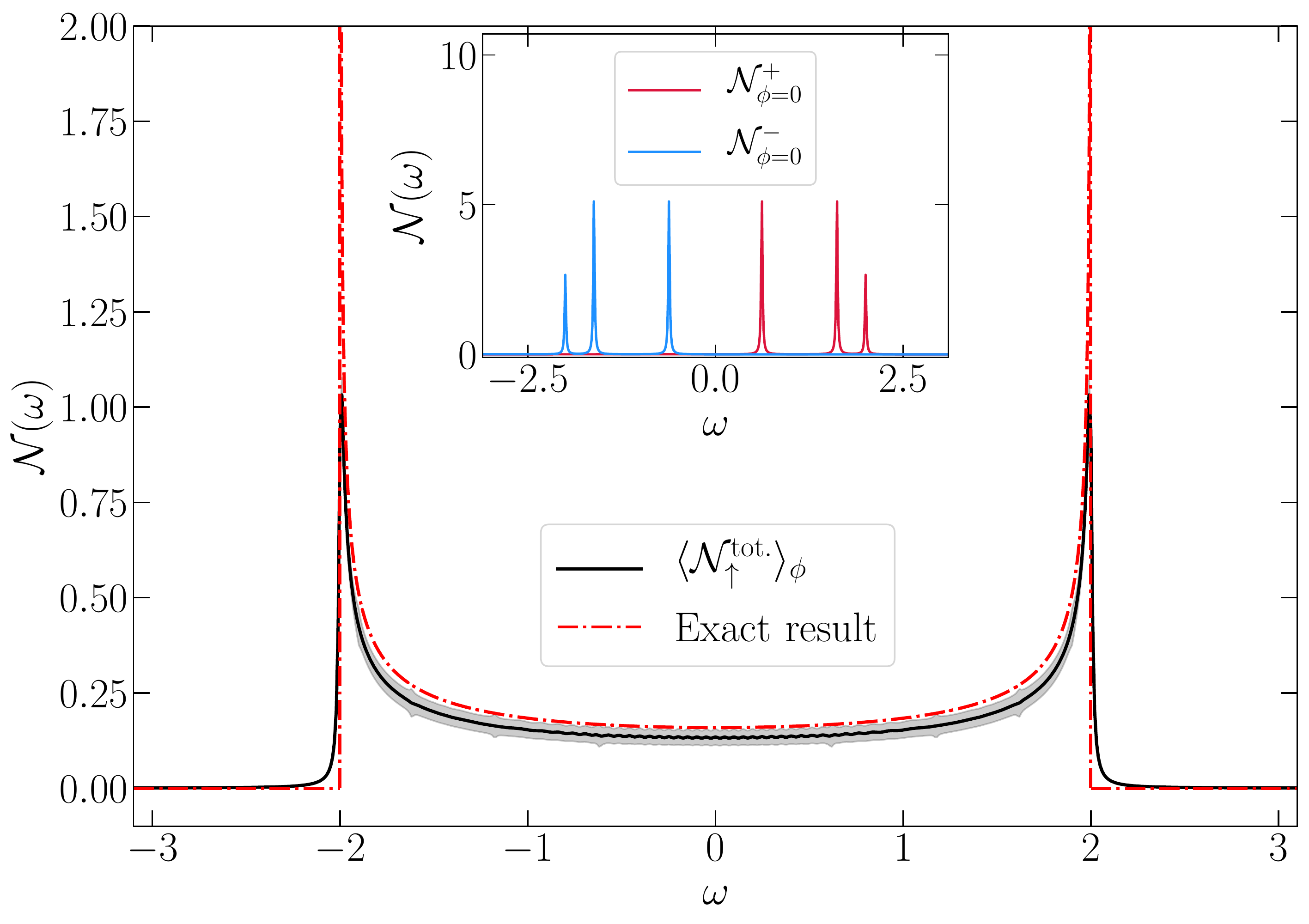}
  \caption{(Color online) Density of states of a non-interacting fermionic chain, with $L=10$, obtained after twisted-boundary averaging and the corresponding exact result in the thermodynamic limit, depicted by the continuous and dashed-dotted lines, respectively. The shading around the former represents the standard error of the mean after the averaging process with 100 phases regularly spaced $\in (0,2\pi]$. The inset shows the numerical result for the case of a single phase, $\phi = 0$, corresponding to the standard periodic boundary conditions, for both the charge and hole excitations.}
  \label{fig:DOS_lx10_U0}
\end{figure}

We argue here that the usage of TBC averaging also helps in obtaining a clear picture of dynamical quantities, in special, of the density of states. To show that, Fig.~\ref{fig:DOS_lx10_U0} displays the single-particle density of states of a non-interacting system with only 10 sites, at half-filling. To start, when considering the simplest case of periodic boundary conditions without twists, $\phi = 0$, the electron excitations ${\cal N}^+(\omega)$ on the Fermi sea at energy $\omega = 0$, correspond to the allowed states of the charge excitations at $k$-points $\pm \frac{3\pi}{5}, \pm \frac{4\pi}{5}$, and $-\pi$, whose energies are $0.618t, 1.618t$, and $2$, respectively. The inset displays the results of the calculated density of states using the Lanczos method. The sequence of three peaks appears at the exact positions given by the previously described energies. Correspondingly, in the case of hole excitations, one is able to have excitations at the originally filled momentum values for the groundstate, $k = 0, \pm \frac{\pi}{5}$, and $\pm \frac{2\pi}{5}$ with energies $-2, -1.618,$ and $-0.618$. Again, these are well captured by the peaks of the ED calculations. The heights in both hole and electron excitations reflect the correspondent multiplictiy of the available energies.

However, these results show how severe are finite-size effects: In the thermodynamic limit, one can integrate the dispersion relation to obtain the analytic form of the density of states as: ${\cal N}_{1d}(\omega) = \frac{1}{2\pi t}\frac{1}{| \sin[2{\cos^{-1}({E/2t})}]|}$, a continuous function in the interval $-2t<\omega<2t$, represented by the dashed-dotted line in Fig.~\ref{fig:DOS_lx10_U0}. Nevertheless, when computing the total density of states for different phases and averaging the results, one is able to closely recover the correct ${\cal N}(\omega)$. This is shown by the continuous line in Fig.~\ref{fig:DOS_lx10_U0}, obtained by averaging one hundred regularly spaced phases in the interval $(0,2\pi]$. The small weights at the energies above and below the maximum and minimum energies, $+2t$ and $-2t$, stem from the functionalization of the delta functions~[{\color{blue} 19}], represented by Lorentzian peaks whose width is given by $\epsilon = 0.01$. We have used this same value of Lorentzian broadening throughout the paper.

Moving from the non-interacting regime, the peaks are naturally broadened by the interactions and a smaller number of phases is sufficient to obtain a converged density of states.

\subsection{Finite-size effects}\label{sec:fse}

\begin{figure}[t]
  \centering
  \includegraphics[width=0.99\columnwidth]{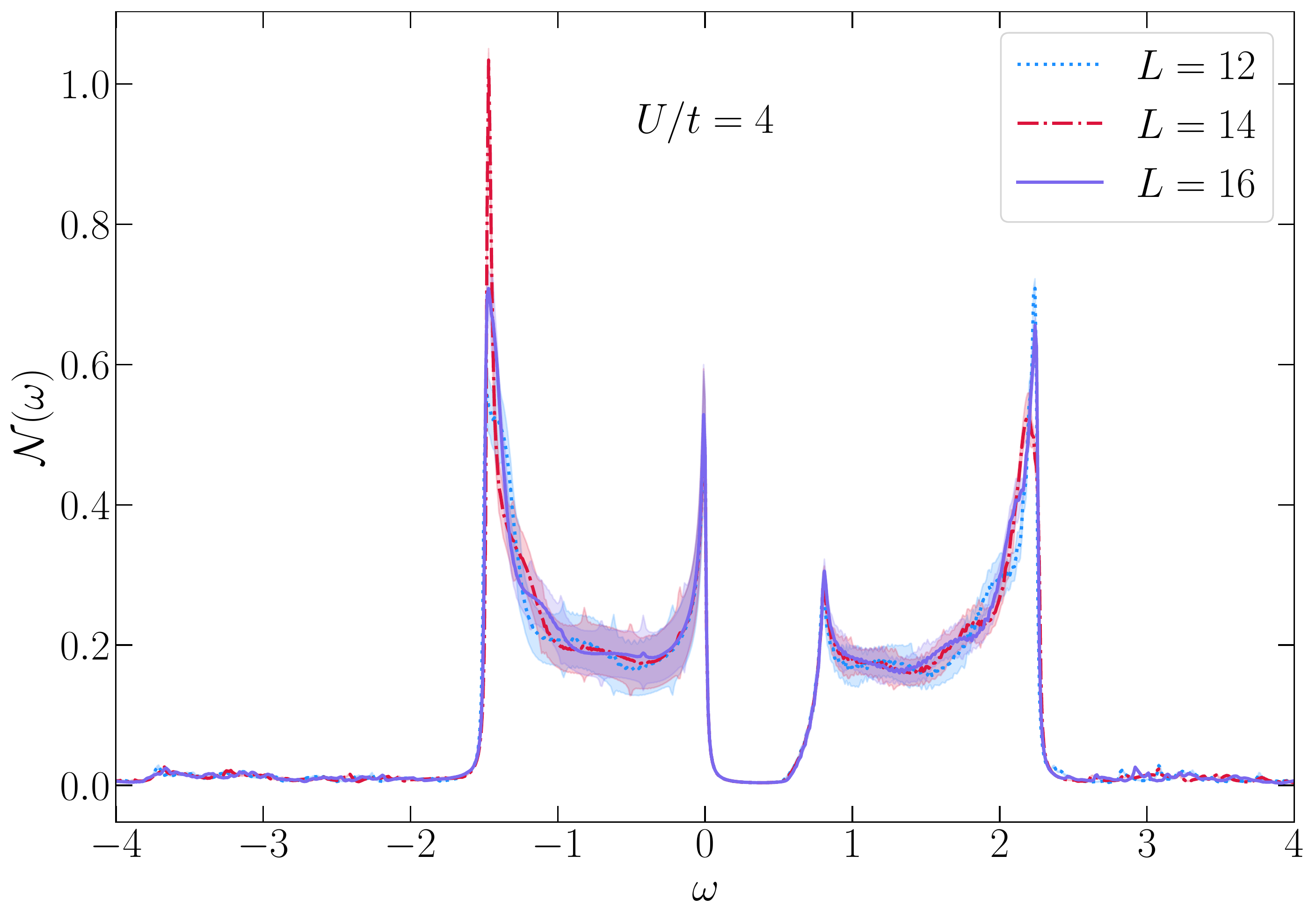}
  \caption{(Color online) Analysis of finite size-effects on the total density of states of the studied superlattice, with a repeated intercalation of repulsive and free sites. These are averaged over 64 phases of the TBC for each system size and the local interaction in the repulsive sites is $U/t = 4$.}
  \label{fig:DOS_finite_size_effects}
\end{figure}

Although some of the finite size effects on the density of states can be removed by employing the TBC averaging, we check in Fig.~\ref{fig:DOS_finite_size_effects} how this convergence to the thermodynamic limit is reached by employing systematically larger superlattices, with a fixed number of phases. The agreement for moderately small system sizes is remarkable. All the qualitative features are already obtained in lattices with $L=12$, in comparison to the ones with $L = 16$, and the small quantitative differences are encompassed by the standard error of the mean that results from the phase averaging, when away from singular energies. This analysis thus confirms the robustness of the TBC averaging in obtaining ${\cal N}(\omega)$ close to the thermodynamic limit results, even for a limited number of boundary conditions used.

\section{Drude weight - homogeneous lattices}
\begin{figure}[t]
  \centering
  \includegraphics[width=0.99\columnwidth]{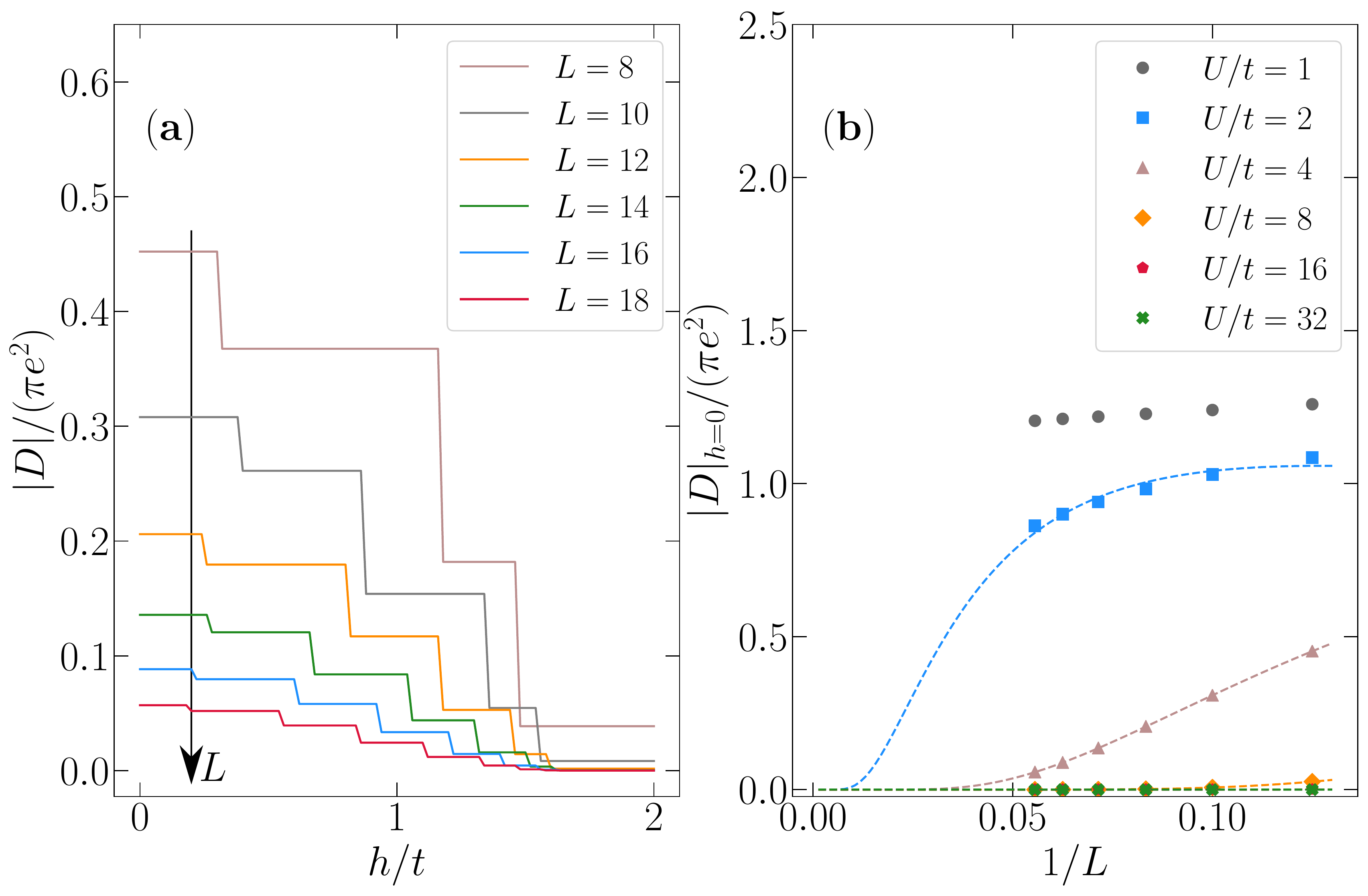}
  \caption{(Color online) In (a), the Drude weight dependence on the Zeeman energy $h$ with $U/t=4$ and different system sizes, whereas in (b), the finite size scaling analysis of the Drude weight for $h=0$. The large finite size effects at small values of $U$ are responsible for the failure of the fitting to the functional form $|D|\propto \sqrt{L/\xi}e^{-L/\xi}$, that arises from the Bethe ansatz equations.}
  \label{fig:Drude_and_DeltaE_vs_hmag_homog}
\end{figure}
For the case of homogeneous lattices, one can obtain the value of the Drude weight via the Bethe ansatz equations~[{\color{blue} 37}]. However, for consistency with the data presented in the main text, we present the Drude weight analysis in Fig.~\ref{fig:Drude_and_DeltaE_vs_hmag_homog}, obtained via Lanczos diagonalization. Unlike for the case of the superlattices, $|D|$ monotonically decreases as a function of the Zeeman energy [Fig.~\ref{fig:Drude_and_DeltaE_vs_hmag_homog}(a)] and it is also smaller as one considers larger system sizes [Fig.~\ref{fig:Drude_and_DeltaE_vs_hmag_homog}(b)]. Therefore, this confirms that the site-dependent interactions, forming the superlattices, are fundamental in inducing a insulator-to-metal transition at half-filling as one applies an external magnetic field on the lattice.

\section{Local density of states}
\begin{figure}[]
  \centering
  \includegraphics[width=0.99\columnwidth]{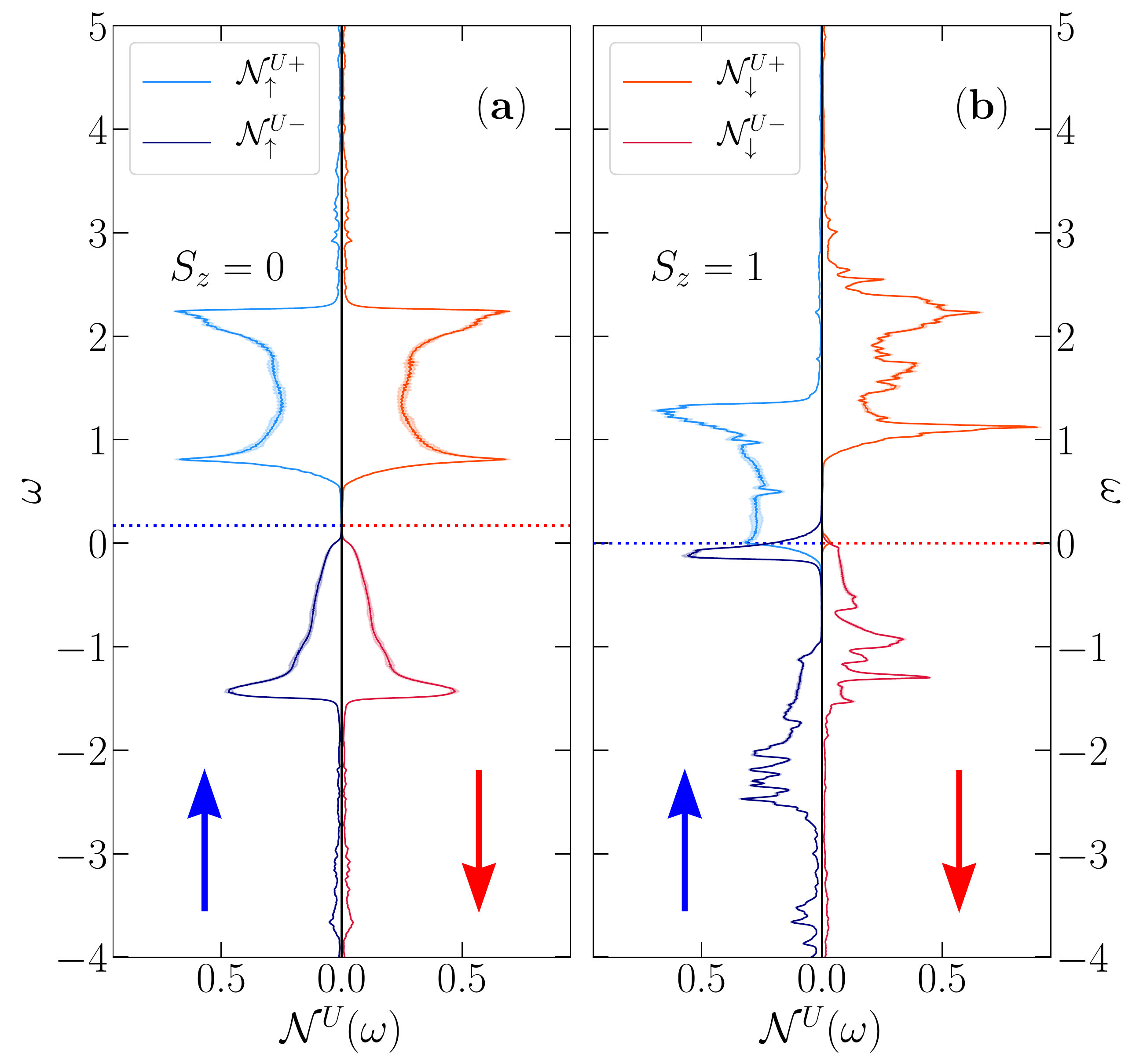}
  \caption{(Color online) The spin resolved local density of states in the repulsive sites (see text for definition) for the cases with zero and finite population imbalances, in (a) and (b), respectively. The dashed lines, are obtained from the integration of the corresponding density of states as to match the correspondent spin-dependent density in the $U>0$ sites; they represent the Fermi energy and are seen to be located at the intersection between the creation and annihilation bands.}
  \label{fig:LDOS_plot}
\end{figure}

The standard argument, used to explain the GMR phenomenon, that in the presence of an external magnetic field the transport is facilitated by the imbalance on the density of states at the Fermi energy between the spin components parallel and anti-parallel to $\vec B$ within magnetic regions, can also be seen in more details in Fig.~\ref{fig:LDOS_plot}. Unlike in the main text, here we separate the \textit{site} contributions to the total density of states, focusing only on the density of states in the repulsive sites, which, accordingly to our simple interpretation, play a role of magnetic regions. This \textit{local density of states} is obtained in a similar manner as the total one, but now, by independently dealing with single-particle excitations in $U>0$ and $U=0$ sites. In other words, we compute it via ${\cal N}^{U}_\sigma \propto \sum_n |\langle \psi_0| \hat c^{U\dagger}_{k,\sigma}| \psi_n^{N_\sigma-1}\rangle|^2 \delta\left(\omega + (E_n^{N_\sigma-1}-E_0)\right) + \sum_n |\langle \psi_0| \hat c^{U}_{k,\sigma}| \psi_n^{N_\sigma+1}\rangle|^2 \delta\left(\omega - (E_n^{N_\sigma+1}-E_0)\right)\equiv {\cal N}^{U+}_\sigma +  {\cal N}^{U-}_\sigma$; $\sigma = \uparrow,\downarrow$, where the (translation invariant) excitations are performed only in the repulsive sites. 

In Fig.~\ref{fig:LDOS_plot}, we compare two cases: when the population is balanced [Fig.~\ref{fig:LDOS_plot}(a)] and when there is a finite imbalance between the two spin components as a result of an external field [Fig.~\ref{fig:LDOS_plot}(b)]. The Mott insulating behavior in the former can still be seen, where now the lower Hubbard band (associated to the single-hole excitations) possess a smaller number of states due to the reduced particle density in repulsive sites, as highlighted in Fig.~2(b) in the main text. However, for finite imbalances [$S_z = 1$ in Fig.~\ref{fig:LDOS_plot}(b)] one can see the dramatic contrast between the local density of states at the Fermi energy: The number of states available for each spin component, which is proportional to the total conductance of this channel, is remarkably different and much smaller for the component anti-parallel to the field. By compiling the difference in the local density of states at the Fermi level for different imbalances of spin populations, we obtain the results in Fig. 1(d) in the main text, for a superlattice with $L=16$.

\section{Obtaining the band insulating transition via an effective two-body model}
In the main text, we argued that there is a saturation field beyond which the system possess a ground state that is fully polarized ($S_z=N_\uparrow/2=L/2$), generating a band insulating behavior. In the absence of interactions, this is easily seen to be related to the field necessary to split the spin $\uparrow$ and $\downarrow$ bands, by overcoming the bandwidth $W=4t$. For finite $U$ values, one can directly compare the lowest energy of two sectors of the Hamiltonian, with $S_z=L/2$ and $S_z=L/2-1$, and investigate what is the necessary field that promotes a crossover in the energies $E(S_z;h)$.

The lowest energy in the polarized case is trivial, since there is only one state in this sector and, thus, $E_0(L/2;h)=-L h$. On the other hand, the ground state energy in the sector where one possess one flipped spin is more involved and is given by $E_0(L/2-1;h)=E_0(L/2-1;h=0)-(L-2)h$. This first term accounts for the lowest energy state at this sector in the absence of an external field. One can easily obtain this energy value, for generic interactions $U$, by transforming the many-body problem onto a two-body problem, when taking into account the degrees of freedom of the hole created in the subspace of $\uparrow$-spins and the additional electron in the $\downarrow$-spins one. By denoting the index of a site where this hole resides by $x_h$ ($x_h=1,2,\ldots,L$), and $x_\downarrow$ ($x_\downarrow=1,2,\ldots,L$) the corresponding site index for the location of the $\downarrow$-spin, one can construct a basis set $|x_h,x_\downarrow\rangle$ to describe the states of the Hamiltonian. If $f(x_h,x_\downarrow)$ represents the amplitude in an eigenstate of the two-body problem where the hole is at site $x_h$ and the $\downarrow$-spin at site $x_\downarrow$, the eigenvalue equation can be written as,

\begin{eqnarray}
 -t[&f&(x_h+1,x_\downarrow)+f(x_h-1,x_\downarrow) \nonumber \\
 +&f&(x_h,x_\downarrow+1)+f(x_h,x_\downarrow-1)] \nonumber \\
 &+&U(1-\delta_{x_h,x_\downarrow})\left(\frac{e^{{\rm i}x_\downarrow\pi}+1}{2}\right)f(x_h,x_\downarrow)=\nonumber \\
 &&E(L/2-1;h=0)f(x_h,x_\downarrow).
\end{eqnarray}

\begin{figure}[b]
  \centering
  \includegraphics[width=0.99\columnwidth]{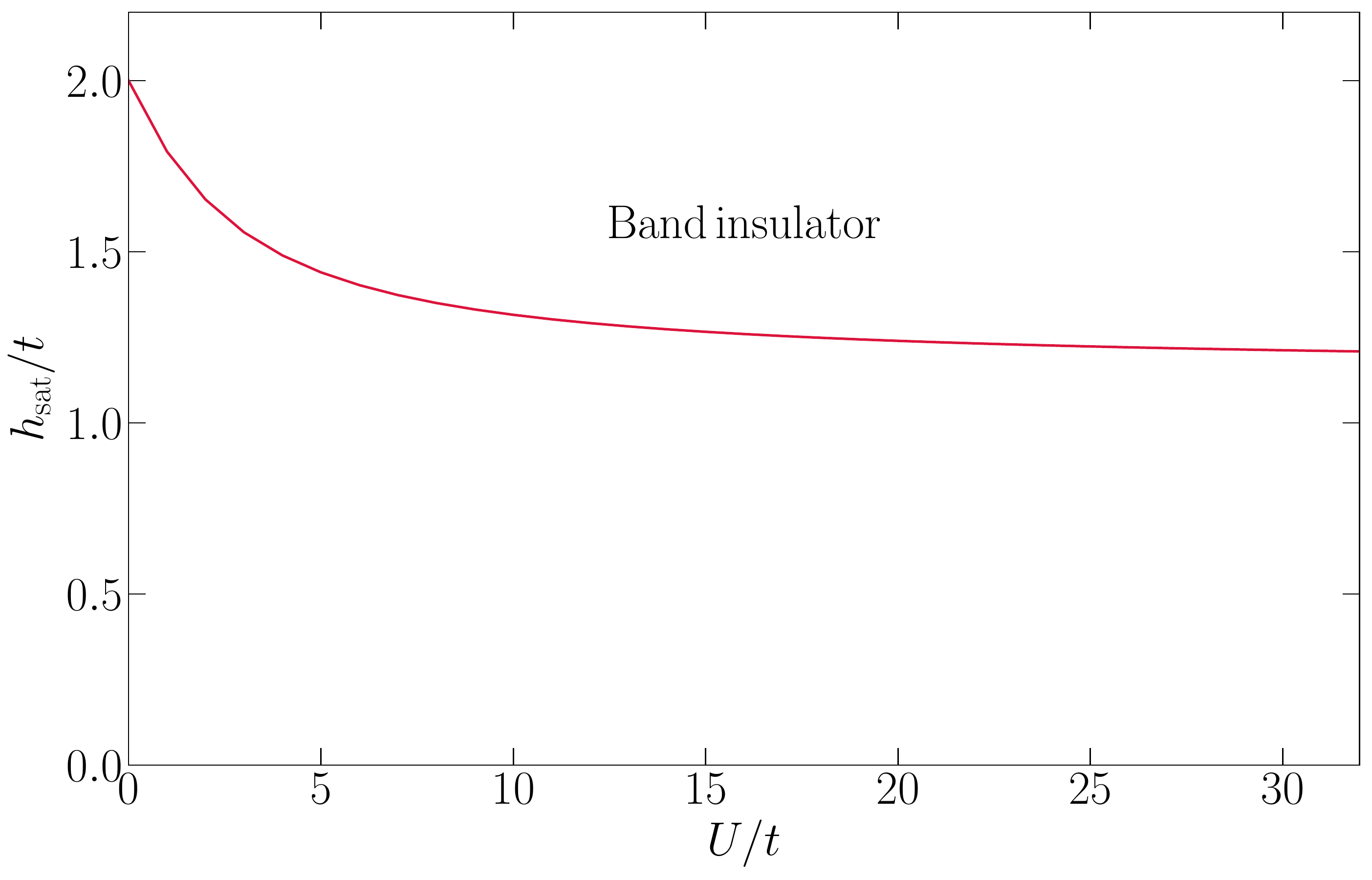}
  \caption{(Color online) Saturation field that marks the onset of the band insulating behavior as a function of the interaction magnitude, in a superlattice with $L=100$, obtained via the analysis of the effective two-body problem.}
  \label{fig:h_sat_effective_model}
\end{figure}

This equation bears similarity with the Bethe ansatz equation~[{\color{blue} 48}], but due to the superlattice structure, it does not admit a standard Bethe ansatz solution. Nevertheless, in matrix form, of dimensions $L^2\times L^2$, it can be easily diagonalized for moderate system sizes $L$. Finally, the critical line $h_{\rm sat}(U)$ is given by when $E_0(L/2;h_{\rm sat}) = E_0(L/2-1;h_{\rm sat})$, which leads to
\begin{equation}
 h_{\rm sat} = -\frac{1}{2}E_0(L/2-1;h=0).
\end{equation}
Figure~\ref{fig:h_sat_effective_model} displays, as an example, this saturation field in a superlattice with $L=100$, in agreement with the result presented in Fig.~5 in the main text.

\end{document}